\documentclass[%
 reprint,
superscriptaddress,
 amsmath,amssymb,
 aps,
prb,
floatfix,
notitlepage,
]{revtex4-1}

\usepackage{txfonts}
\usepackage{graphicx}
\usepackage{color}%
\usepackage[caption=false]{subfig}

\newcommand{\beq}{\begin{equation}}
\newcommand{\eeq}{\end{equation}}
\newcommand*\diff{\mathop{}\!\mathrm{d}}
\def\half{\frac{1}{2}}
\def\br{{\bf r}}
\def\dd{{\diff}}
\def\dr{{\!\!\dd\br\,}}
\def\calF{\mathcal{F}}
\def\calS{\mathcal{S}}
\def\calT{\mathcal{T}}

\def\occ{\mathrm{occ}}

\def\H{{\mathrm{H}}}
\def\s{{\mathrm{s}}}
\def\t{{\mathrm{t}}}
\def\xc{{\mathrm{xc}}}
\def\TF{{\mathrm{TF}}}
\def\vW{{\mathrm{vW}}}
\def\SGA{\mathrm{SGA}}

\def\GGA{\mathrm{GGA}}
\def\D{{\mathrm{D}}}
\def\T{{\mathrm{T}}}

\begin{document}

\title{Towards accurate orbital-free simulations: a generalized gradient approximation for the non-interacting free energy density functional}
\author{K.~Luo}
\email{kluo@carnegiescience.edu}
\affiliation{Geophysical Laboratory, Carnegie Institution,
 5251 Broad Branch Road NW, Washington D.C. 20015}
\author{V.V.~Karasiev}
\email{vkarasev@lle.rochester.edu}
\affiliation{Laboratory for Laser Energetics, University of Rochester,
 Rochester, NY 14623}
\author{S.B.~Trickey}
\email{trickey@qtp.ufl.edu}
\affiliation{Quantum Theory Project, Department of Physics and 
Department of Chemistry, University of Florida, Gainesville, FL 32611}

\date{submitted Dec.\ 04, 2019, revised Jan.\ 27, 2020}

\begin{abstract}
For orbital-free {\it ab initio} molecular dynamics, especially on systems in 
extreme thermodynamic conditions, we provide the first
pseudo-potential-adapted generalized gradient approximation (GGA)
functional for the non-interacting free energy.  This is achieved by 
systematic finite-temperature extension of our recent LKT ground state 
non-interacting kinetic energy GGA 
functional (Phys. Rev. B \textbf{98}, 041111(R) (2018)).
We test the performance of the new functional first via static lattice 
calculations on crystalline aluminum and silicon.
Then we compare deuterium equation of state results against both 
path-integral Monte Carlo and conventional (orbital-dependent) 
Kohn-Sham results.
The new functional, denoted LKTF, outperforms the previous best semi-local free energy functional, VT84F (Phys.\ Rev.\ B \textbf{88}, 
161108(R) (2013)), and provides modestly faster simulations.
We also discuss subtleties of identification of kinetic and entropic 
contributions to non-interacting free-energy functionals obtained by 
extension from ground state orbital-free kinetic energy functionals.  
\end{abstract}


\maketitle

\section{Context and Motivation}
\label{sec:contextmotive}

Warm dense matter (WDM) has been a research topic of substantial
recent interest because of its 
importance in high energy density sciences
and its inherently quantum mechanical nature \cite{Graziani2014Book}. 
WDM has been a challenge both experimentally and theoretically. 
Despite progress, it remains so.
Though advances in experimental facilities and techniques are making parts of 
the relevant state space accessible, the value and urgency of reliable, 
computationally affordable theoretical methods still is undeniable.  
However, conventional methods are unaffordable for application over the 
entirety of the typical temperature range of interest. For example, 
path-integral Monte Carlo (PIMC) takes advantage of the Trotter expansion 
at very high temperatures but becomes intractable lower into the WDM 
regime. Conversely, stochastic  density
functional theory (DFT) remains computationally expensive at low temperature \cite{Baer2013PRL,Baer2018PRB}.

For ordinary condensed matter conditions, ground state 
DFT \cite{Hohenberg1964PR} in its conventional
Kohn-Sham (KS) realization \cite{KohnSham1965PR} has achieved enormous
success.  That is thanks to the elegant balance between computational
cost and accuracy provided by KS DFT.  By extension, the {\it de
  facto} standard methodology for WDM, e.g., for prediction of
equations of state, is {\it ab initio} molecular dynamics (AIMD) with
forces from Mermin free-energy DFT \cite{Mermin1965PR}.

For general state conditions, however, the conventional KS
implementation of FT-DFT, with its explicit orbital dependence in the
form of solution of the KS equations, scales computationally no better
than $N_{\occ}^3$, with $N_{\occ}$ the number of occupied KS orbitals.
 For gapped systems, locality or sparsity can be exploited to achieve 
linear scaling \cite{Nakata2017} but this approach lacks the generality
of applicable state conditions essential for WDM.
As the electron temperature (and/or system size) grows, at some point
KS-AIMD calculations become impractical (unaffordable) because of the
enormous number of non-negligibly occupied KS states. Orbital-free
molecular dynamics (OFMD) is an attractive alternative because its
computational cost scales linearly with system size irrespective of
the particular system state.

With recent advances in approximate non-interacting kinetic density
functionals $T_{\s}$, ground-state OFMD is beginning to be a viable
alternative to low-T KS-AIMD. 
Both semi-local and non-local functionals have achieved mixed successe 
in treating condensed phases and their ingredient atoms, molecules, and 
clusters and solids.  Such functionals are either constraint-based
and non-empirical \cite{LevyOuYang1988, Karasiev2012PRB, Luo2018PRB, Luo2018CPL,  Constantin2018JPCL, MGP2018JCP, Mi2019PRB,Lehtomaki2019PRB, Witt2019PRBa,Witt2019PRBb, Witt2019JCP} or semi-empirical \cite{Constantin2017,Constantin2018PRB}.
With any significant ground-state advance, an 
obvious, important associated step is generalization to a non-interacting free
energy functional $\calF_{\s}$.  In this work, we make that step based
upon a recently proposed ground state $T_{\s}$ functional, LKT
\cite{Luo2018PRB}.  It has the novel property of being adapted
specifically to working with pseudo-densities, such as almost always
are used in AIMD calculations.  Thus LKT satisfies known constraints
on $\calF_{\s}$ for pseudo-densities, not physical densities.  Hence LKT
is non-universal by construction to achieve good performance from a
semi-local functional. But it is \textit{not empirical}.

The next section summarizes free energy DFT to establish notation, 
conventions, and correspondence with ground state KS-DFT. It 
then summarizes the T-dependent dimensionless gradient variables
developed in Ref.\ [\onlinecite{Karasiev2012PRB}] and uses them to
generalize the LKT $T_{\s}$ to $\calF_{\s}$.  Section \ref{sec:computdetails} 
summarizes matters of computational technique, after which 
Section \ref{sec:results} presents calculated results and comparisons.
We conclude with discussion and summary in Section \ref{sec:discussion}.

\section{Free energy density functionals}
\label{sec:freeenergy}

In the grand canonical ensemble, the electronic grand potential
$\Omega$ for a system of average electron number $\langle N \rangle$
under external potential $v$ is minimized by the equilibrium
electronic number density $n_{\mathrm eq}$, that is, there is a one-to-one
mapping between $v$ and $n_{\mathrm eq}$ (see
Ref.~[\onlinecite{Mermin1965PR}] for details).  The 
electronic grand potential can be written as a density functional 
\begin{equation}
\Omega [n, \T] = \calF[n,\T] +\int \dr (v(\br) - \mu ) n(\br)\,,
\label{eq:grandpotential}
\end{equation}
where $\mu$ and  $\T$ are the chemical potential and electronic 
system temperature. 
The universal free energy functional $\calF[n,\T]$ can
be constructed formally by constrained search.  As in the 
ground-state KS scheme, the free energy functional is decomposed into 
three pieces,
\begin{equation}
\calF[n,\T] = \calF_{\s} [n,\T] +\calF_{\H}[n] + \calF_{\xc}[n,\T]\,,
\label{eq:calFdefn}
\end{equation}
where $\calF_{\s}$, $\calF_{\H}$, and $\calF_{\xc}$ are the non-interacting 
free energy, the classical Coulomb free energy (or Hartree energy), and 
the exchange-correlation (XC) free energy, respectively.
$\calF_{\H}$ has a simple, explicit density dependence, hence needs no
attention. 

In the conventional
use of the KS decomposition,  the non-interacting free energy 
$\calF_{\s} = \calT_{\s} - \T\calS_{\s}$,  is 
treated exactly, with the orbital-dependent, non-interacting KE 
and entropy given by
\vspace*{-3pt}
\begin{equation}
\calT_{\s} [n,\T] = -\half \sum_{j=1}^{N_{\occ}} \int \dr  f_j 
 \varphi_j^* (\br) \nabla^2 \varphi_j(\br) 
\end{equation}
and 
\begin{equation}
\calS_{\s} [n,\T] = -k_{\rm B} \sum_{j=1}^{N_{\occ}} \left[f_j \ln f_j + (1-f_j) \ln (1-f_j) \right]\, .
\end{equation}
Here $\varphi_j$ are thermally occupied KS orbitals with $j=1, \dots N_{\occ}$. 
The Fermi-Dirac distribution function is 
$f_j 
= 1/(1+e^{\beta (\varepsilon_j - \mu)})$
where $\varepsilon_j$ is $j$th eigenvalue of the KS equation and 
$\beta = 1/(k_{\mathrm B} \T)$ is the inverse temperature with Boltzmann constant 
$k_{\mathrm B}$. In computational practice, the chemical potential $\mu$ 
is determined via $\sum_{j=1}^{N_\occ} f_j = N$, the number of electrons.  

In this context, the only approximation needed is for the XC free
energy $\calF_{\xc}[n, \T]$.  There has been recent progress on both
local density approximations (based on the homogeneous electron gas,
HEG) in Refs.~[\onlinecite{KSDT2014PRL,Groth2017PRL,KTD2019PRB}] and on
a generalized gradient approximation \cite{KDT2018PRL} for $\calF_{\xc}[n, \T]$.

Solution of the conventional KS eigenvalue problem requires diagonalization 
or equivalent.  That is the source of the computational cost scaling 
no better than $N_{\occ}^3$ already noted. Such scaling poses a major obstacle to 
routine WDM simulation, as already remarked.
Orbital-free DFT (OFDFT) offers the potential to remove this barrier.

\subsection{Generalized gradient approximations}

Two approximate functionals are required in free-energy OFDFT, $\calF_{\s}$ and
$\calF_{\xc}$.  Our focus is on the first.  
   
The most widely used, though far from optimal $\calF_{\s}$ approximation 
in free-energy OFDFT is the Thomas-Fermi (TF) functional \cite{Feynman1949PR}. 
By making a local density approximation based on the HEG as paradigm, 
evaluation of Eq.~(\ref{eq:grandpotential}), $\Omega^{\mathrm{HEG}}$, leads to 
the TF approximate free energy 
\begin{equation}
\calF^{\TF}_{\s} [n, \T] = \int \dr f_{\s} ^{\TF}(n, \T)\,,
\end{equation}
and associated free energy density 
\begin{equation}
f_{\s} ^{\TF}(n, \T) = \frac{\sqrt{2}}{\pi^2 \beta^{5/2}}\left[-\frac{2}{3}I_{3/2}(\beta \mu) + \beta \mu I_{1/2}(\beta \mu)\right] \,.
\label{eq:feTF}
\end{equation}
(Note that free energy densities are unique only up to a gauge transformation;
here and throughout we use conventional forms.)
The Fermi-Dirac integrals \cite{Blakemore1982,BartelBrackDurand1985} are
\begin{equation}
I_{\alpha} (\eta) \equiv \int_{0}^{\infty}  \frac{x^\alpha}{1+e^{x-\eta}} \dd x\,.
\end{equation}
The chemical potential $\mu$ can be determined from
\begin{equation}
n=-\frac{1}{V} \left. \frac{\partial \Omega^{\mathrm{HEG}}}{\partial \mu}\right|_{T,V} = \frac{\sqrt{2}}{\pi^2 \beta^{3/2}} I_{1/2}(\beta \mu).
\end{equation}

In terms of the reduced temperature 
\begin{equation}
t=\T/\T_F = \frac{2}{\beta[3\pi^2 n]^{2/3}} \; 
\end{equation}
$ I_{1/2}(\beta \mu) =n \pi^2 \beta^{3/2} / \sqrt{2} 
=2t^{-3/2}/3$ and Eq. \eqref{eq:feTF} becomes
\begin{equation}
f_{\rm s} ^{\rm TF}(n, \T) = \tau_0^{\rm TF}(n) \kappa(t)
\end{equation} 
with 
\begin{equation}
\tau_0^{\rm TF}(n) = \frac{3}{10} (3\pi^2)^{2/3} n^{5/3} 
\end{equation}
and 
\begin{equation}
\kappa(t) = \frac{5}{2} t^{5/2} \left[-\frac{2}{3}I_{3/2}(\beta \mu) + \beta \mu I_{1/2}(\beta \mu)\right]\,.
\end{equation}

Beyond the HEG, the second-order gradient approximation (SGA) 
for the non-interacting free-energy density is 
\begin{equation}
f_{\s}^{\SGA}(n, \nabla n, \T) = f_{\s} ^{\TF}(n, \T)  + 8 h(t) \frac{|\nabla n|^2}{8n}\,,
\end{equation}
with
\begin{equation}
h(t) = -\frac{1}{24} \frac{I_{1/2}(\beta \mu) I_{-3/2}(\beta \mu)}{I_{-1/2}^2(\beta \mu)} \,.
\end{equation}
It is convenient to use $\tilde{h} = 72 h$ because 
 $\lim_{t\to 0} \tilde{h}(t) =1$. 

The well-documented limitations of the SGA motivate generalized gradient
approximations (GGAs).    
Some time ago, a systematic means of promoting a ground-state 
GGA non-interacting functional to 
become a non-interacting free energy GGA was put forth \cite{Karasiev2012PRB}.
Ground-state functionals are expressed as a function of the 
dimensionless reduced density gradient
\begin{equation}
s(n,\nabla n) :=  \frac{|\nabla n|}{(2k_F)n}
 =  \frac{1}{2(3\pi^2)^{1/3}} \, \frac{|\nabla n|}{n^{4/3}}\,,
\label{sdefn}
\end{equation}
By examination of the finite-T gradient expansion,  
Ref.~[\onlinecite{Karasiev2012PRB}] identified the
proper finite-T reduced density gradients for the kinetic
and entropic contributions, to wit,
\begin{eqnarray}
s_{\tau} (n,\nabla n, \T) &=& s(n,\nabla n) \sqrt{\frac{\tilde{h}(t) - t \tilde{h}'(t)}{\xi(t)}} \label{stau}
\\
s_{\sigma} (n,\nabla n, \T) &=& s(n,\nabla n) \sqrt{\frac{t \tilde{h}'(t)}{\zeta(t)}} \;.  \label{ssigma}
\end{eqnarray}
Here the $t$-dependent functions are 
\begin{align}
\xi (t) &= \kappa(t) - t \kappa'(t) \,,
\\
\zeta(t) &= - t \kappa'(t)\, ,
\end{align} 
and primes denote differentiation with respect to the indicated variable.
The finite-temperature GGA free energy functional then has a kinetic
and entropic term, 
\begin{equation}
\calF_{\s}^{\GGA}[n,\T] = \int \dr\, \tau_0^{\TF}  \left[ \xi(t) F_{\tau}(s_{\tau}) - \zeta(t) F_{\sigma}(s_{\sigma}) \right]  \,,
\end{equation}
with distinct enhancement factors, $F_{\tau}$ and $F_{\sigma}$.

Evidently, the zero-T GGA enhancement factor is only for the kinetic energy, 
that is $F_{\tau}(s_\tau) \rightarrow F_t(s)$.  
In addition, therefore, to the replacement $s\rightarrow s_\tau$,
the entropic enhancement factor $F_{\sigma}$ must be constructed. 
A thermodynamic Maxwell
relation relates the two exactly but the resulting differential equation is
not trivial to solve \cite{Karasiev2012PRB}.   An 
identity for the SGA  \cite{Karasiev2012PRB} 
\begin{equation}
F_{\sigma}(s_\sigma) = 2 - F_{\tau}(s_\sigma)\,
\label{twominusF}
\end{equation}
is a useful approximation for GGA construction.  
To date it has proven reasonably successful.
  For instance, VT84F, an earlier GGA free energy functional, used 
(\ref{twominusF}) to yield reasonably good performance in the WDM regime \cite{Karasiev2013PRB}. Detailed numerical assessment of Eq.\ (\ref{twominusF}) in the 
present case is 
given in the Supplemental Information \cite{SuppMat}.

For clarity of analysis, we include the ground-state approximate 
functionals $\TF\lambda \vW$, with $\lambda=1/5$ or $1/9$. Their
 enhancement factor is
\begin{equation}
F_{\t}^{\TF\lambda \vW}(s) = 1+\lambda \frac{5}{3} s^2 \,.
\label{eq:tflambdavw}
\end{equation}
Here ``vW'' denotes the von Weizs\"acker KE functional.
We note that such TF plus scaled vW functionals with $\lambda < 1$
violate the positivity requirements on 
the Pauli potential $v_\theta$ that is the functional derivative of the
Pauli KE ${\mathrm T}_\theta$ in the rigorous 
decomposition \cite{LevyOuYang1988} 
\beq
T_{\s} = T_{\vW} + T_\theta, \;\; T_\theta \ge 0 \; . 
\label{eq:Tthetadcomp}
\eeq
 Nonetheless there is a literature of using 
TF$\frac{1}{5}$vW for the ground state, hence it is a useful context to 
assess its performance when extended to finite $T$.
Note also that TF$\frac{1}{9}$vW is the Perrot functional \cite{Perrot1979PRA},  

\subsection{Adaptation to pseudo-densities}

The aforementioned exact positivity conditions for the ground-state 
KE functional are $T_\theta \ge 0$ and $\delta T_\theta / \delta n \ge 0 \, \forall \mathbf r$.  
These are powerful tools for constraint-based, non-empirical development 
of ground-state approximate functionals.  In particular,
the ground-state limit of the VT84F functional \cite{Karasiev2013PRB}
was developed to meet those constraints (as well as others) for realistic
atomic densities.  Such densities have cusps at the nuclei \cite{Kato1957}.  
VT84F therefore is non-universal in the particular sense in which ``universal''
is used in DFT. VT84F was adapted, by construction, to properties of the
densities characteristic of bare Coulomb external potentials.  

By design, the pseudo-densities almost always used in AIMD calculations do not
have such Coulombic cusps.  Instead they have zero gradients at the
origin.  In that computational setting, VT84F (at T=0 K) can perform
unreliably.  Our response was to put forth the LKT ground-state
functional \cite{Luo2018PRB}.  It was formulated specifically to meet
the rigorous positivity constraints in conjunction with ordinary pseudo-densities.

In the present work, we use the free-energy GGA methodology
\cite{Karasiev2012PRB} just summarized to promote LKT
\cite{Luo2018PRB} into a free energy density functional,
``LKTF''. The LKT enhancement factor is
\begin{equation}
F_{\theta}^{\mathrm{LKT}} (s) =1/\cosh (a s) \;\; \textrm{with} \;\; a=1.3 \; .
\label{LKTenh}
\end{equation}
Specifically, we have
used the variables in Eqs.\ (\ref{stau}), (\ref{ssigma}) and the 
approximate relationship Eq.\ (\ref{twominusF}) between $F_\tau$
and $F_\sigma$.   

\section{Computational Details}
\label{sec:computdetails}
The calculations were of two types.  One is electronic free energy
minimization in the field of static ions (``static lattice'').  The
other is AIMD.  All the calculations used the ground-state
Perdew-Zunger local density approximation for the XC free energy
functional \cite{PerdewZunger1981PRB} without explicit temperature
dependence. This choice (the ground-state approximation) is for
clarity of comparison among non-interacting functionals.  In
calculations for actual materials properties, proper free-energy XC
functionals should be used
\cite{KDT2018PRL,KarasievCalderinTrickey2016}.

The static lattice OF calculations were done using a locally modified 
version of the {\sc profess} \cite{PROFESS3.0} 
code with finite-temperature 
capability.  Comparison finite-T
 KS calculations were done with {\sc abinit} version 8.8 \cite{Gonze2016CPC}.
We chose two representative simple elements, face-centered cubic (\textrm{fcc}) Al 
and cubic diamond (\textrm{cd}) Si. 
Both conventional KS and OF calculations used 
the BLPS \cite{Huang2008PCCP} local
pseudo-potential. The KS calculations used plane wave energy 
cutoffs of 800 eV and 850 eV for Al and Si respectively. 
Monkhorst-Pack k-point sampling convergence was used with 
4 atoms in fcc symmetry with a $15\times 15\times 15$ grid and 8 atoms in 
\textit{cd} symmetry with a $9\times 9\times 9$ grid.
Temperatures were from 1 to 10 eV in 1 eV increments.
For Al, the bulk density range was 2.3 to 3.3 g/cm$^3$ sampled at 
0.2 g/cm$^3$ intervals. The corresponding values for Si were 2.0 to
 2.6 g/cm$^3$ at 0.1 g/cm$^3$ intervals. All bands with occupation 
$\ge 10^{-6}$ were included.
The resulting number of bands used is listed in the Supplemental Materials
\cite{SuppMat}.

OF calculations were done with a representative group of one-point 
non-interacting free 
energy density functionals: TF, Perrot (i.e., TF$\frac{1}{9}$vW), TF$\frac{1}{5}$vW, VT84F, and LKTF. The TF$\lambda$vW forms were implemented via the
finite-T methodology summarized above and with Eq.\ (\ref{twominusF}),
which is exact for those forms. In addition, we include a relatively recently
developed non-local (two-point) non-interacting functional which has
had some success\cite{Sjostrom2014PRL}. We denote it as
SD$\beta$-vW14F.  

The AIMD calculations were for the equation of state (EOS) of hydrogen (H), 
deuterium (D), and Al. Whether driven by conventional KS or OFDFT
forces, the calculations were performed on the same footing with the {\sc
  profess}@{\sc Quantum-Espresso} package \cite{Karasiev2014CPC} 
and the same ground-state XC functional (PZ) as in the static cases. 
The bulk densities used were chosen such that the D EOS results could 
be compared with published PIMC values \cite{Hu2011PRB}. 
For H and D, in both the KS-AIMD and OF-AIMD calculations the electron-ion 
interaction was treated via a deep local pseudopotential 
\cite{Karasiev2014CPC} with core radius 0.25 bohr. For Al, the KS-AIMD
calculations used the non-local PAW dataset (Al.pz-n-kjpaw\_psl.0.1.UPF)
\cite{DalCorso2014CPC},
and the PZ XC functional, while the OF-AIMD calculations used the 
aforementioned BLPS. 

All the orbital-free calculations used a real-space grid size of
$64^3$ or $96^3$ for H(D)
and $128^3$ for Al depending on the bulk
densities. 
The number of atoms was 108 for H and D and 128 for Al.  The
time step varied from 0.0126 fs to 0.357 fs. $\Gamma$ point sampling
was used for the KS-AIMD unless stated otherwise.  Ion temperatures
were regulated by Andersen thermostat.  After equilibration, each system was run for 2000 steps.
Pressures were averaged over those 2000 steps, yielding a maximum
standard deviation relative to the average pressure of 5\%. 

\begin{figure}[!htbp]
	\centering
	\includegraphics[width=\linewidth]{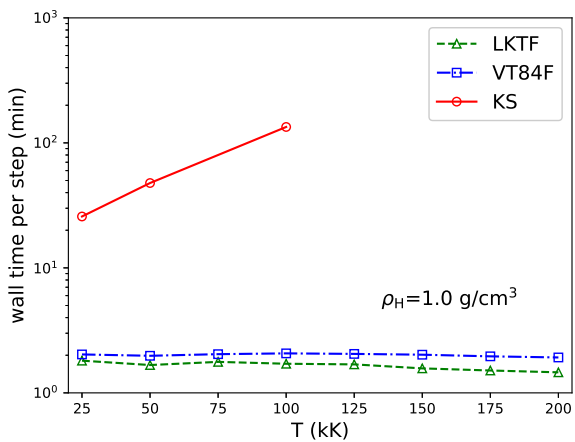}
	\caption[TimeReport]{Average WALL time per MD step as a function of T for ordinary KS-AIMD, and OF-AIMD with VT84F, and LKTF functionals. Hydrogen density is $\rho_{\H}$ = 1.0 g/cm$^3$. The KS cost grows while 
the LKTF and VT84F cost per step is T-independent. As noted before, at  
T = 0 K, LKT SCF convergence is faster than VT84F. 
	}
	\label{fig:timereport}
	\vspace*{-8pt}
\end{figure}

\section{Results}
\label{sec:results}

\subsection{Computational Cost}
\label{sub:computationalcost}

First, we consider the actual computational cost of OF-AIMD against
KS-AIMD for H.  For both types, the time per
step was averaged over 6000 steps.  The computations were performed on
Intel E5-2698v3 processors with 4 GB of RAM per core.  KS-AIMD used two 
nodes, while OF-AIMD used one. Each node comprised 32 cores.  The
wall time per step in units of minutes is shown in
Fig.~\ref{fig:timereport}.  The KS-AIMD cost actually grows exponentially,
while the time per step of all the OF calculations (LKTF, VT84F, TF)
is ${\mathrm{T}}$-independent.  As expected, TF (not shown) runs fastest, 
a consequence of its simple locality.  Typically LKTF
requires fewer iterations to reach its converged electron 
density than the other semi-local functional, VT84F. That advantage
is reflected in the WALL time.  A slight 
decrease in WALL time is observed for both LKTF and VT84F as T grows.
We surmise that this is a consequence of growing homogeneity of the
electron distribution as ${\mathrm{T}}$ increases but have not investigated.

\subsection{Static lattice EOS}
\label{sub:staticcurves}
The main focus of this work is to make the free-energy generalization
of LKT and to explore its direct consequences.  Improvements due to 
making alternative choices of ground-state kinetic energy density 
functionals, refined choice of XC 
functional, or alternative pseudo-potential forms are outside the
scope of the present report. Thus, for comparison we select a  
representative but clearly non-exhaustive set of 
kinetic energy density functionals.
\subsubsection{fcc Al}

\begin{figure}[]
        \centering
        \subfloat[2a][isochoric]{
        \centering
	\includegraphics[width=0.9\linewidth]{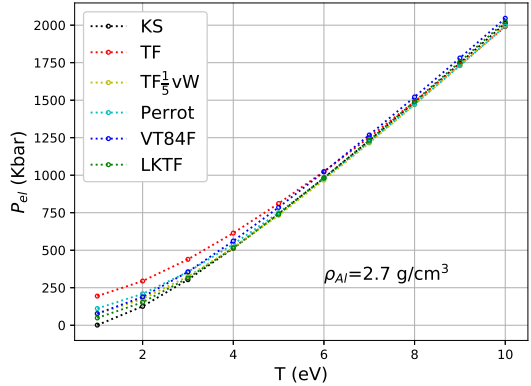}
         \label{fig:Al-rho-2.7}
        }
        \\
        \subfloat[2b][isothermal]{
        \centering
	\includegraphics[width=0.9\linewidth]{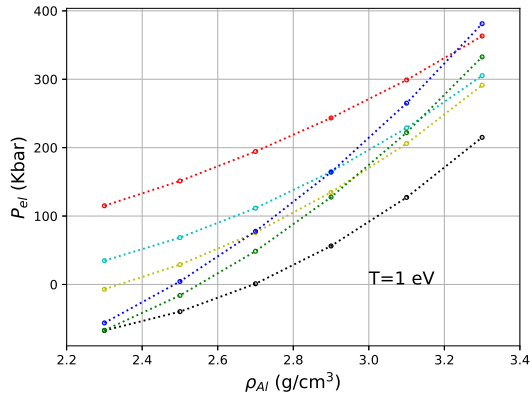}
       \label{fig:Al-T-1eV}
        }
 
    	\caption[Al]
	{ Static lattice fcc Al electronic pressures from various OF functionals 
compared with conventional KS calculations. 
    Top panel: Pressure as function of T for fixed material density 
$\rho=2.7$ g/cm$^3$;
    Bottom panel: Isothermal pressure (T = 1 eV) as function of 
material density.
}
	\vspace*{-8pt}
\end{figure}

As a representative case, for fcc Al we compared the electronic
pressures of various OF functionals against those from the KS
reference calculations.  Fig.~\ref{fig:Al-rho-2.7} shows the results
for bulk density $\rho=2.7$ g/cm$^3$.  Across the entire temperature
range, of all the OF functionals LKTF stays closest to the KS
data. At low temperatures, however, the OF functionals fail to
reproduce the conventional KS results.  To assess the performance of LKTF for
slightly higher pressure and temperature, we analyzed the isothermal
pressure at T = 1  eV for 2.3 $\le \rho \le 3.3$g/cm$^3$. See
Fig.~\ref{fig:Al-T-1eV}.  From 2.2 to 2.9 g/cm$^3$, LKTF values
remain closest to the conventional KS data, but for higher densities
$\TF\frac{1}{5}\vW$ is slightly better.  Except for LKTF at the
lowest density, none of the OF functionals does very well in this
comparison. 

All the data for this section, both for fixed $\rho$ and fixed T, 
are included in the Supplemental Material.

%
%

\subsubsection{cd Si}

\begin{figure}[]
        \centering
        \subfloat[2a][isochoric]{
        \centering
        	\includegraphics[width=0.9\linewidth]{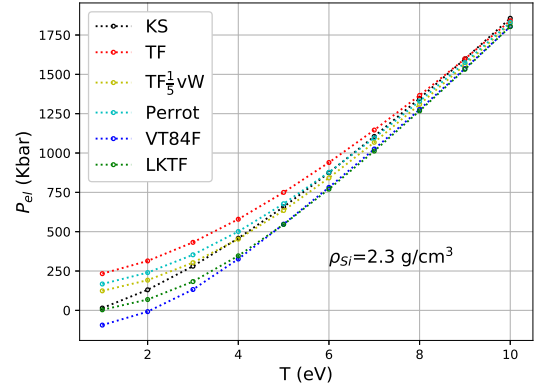}
    \label{fig:Si-rho-2.3}
        }
        \\
        \subfloat[2b][isothermal]{
        \centering
        	\includegraphics[width=0.9\linewidth]{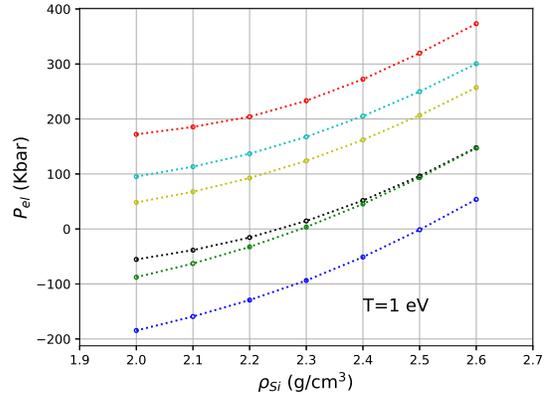}
       \label{fig:Si-T-1eV}
        }
 
	\caption[Si]
	{ 
Electronic pressure prediction comparison for various OF functionals 
compared with conventional KS results for static lattice cubic diamond (cd) Si. 
	Top panel: Pressure as function of T for fixed material density 
$\rho=2.3$ g/cm$^3$;
    Bottom panel: Isothermal pressure (T = 1 eV) as function of material density.
	}
	\vspace*{-8pt}
\end{figure}

Fig.\ \ref{fig:Si-rho-2.3} shows the electronic pressures for \textrm{cd} Si
at $\rho=2.3$ g/cm$^3$, close to the ambient bulk density. At T = 1 eV,
among all the OF functionals, the LKTF pressure is almost identical 
to that from the conventional KS reference.  However, as T grows, the LKTF 
EOS tends toward the VT84F EOS and the two are indistinguishable above
$\T \approx 4$eV.  Both lie below the conventional KS EOS. 
Whether this behavior is a shared flaw of the parent ground-state
GGAs or is a sign of some limitation of the 
finite-T extension of the reduced gradient variable 
(summarized above) or some combination is unclear.  
In contrast,  TF$\lambda$vW approaches the KS EOS above   
$\T \approx 3$ eV, with the choice of
$\lambda=\frac{1}{5}$ outperforming $\lambda=\frac{1}{9}$ and
$\lambda=0$. Note however, that TF$\frac{1}{5}$vW goes a bit below
the conventional KS pressures above about T = 5 eV. Eventually, of 
course, everything goes to TF ($\lambda=0$).
 
To gain understanding of these observations, we
used the thermodynamic relation
\begin{equation}
P_{el} = -\left. \frac{\partial \calF_{el}}{\partial V}\right|_{\T,N} \,
\end{equation} 
to compute the pressure contributions from the internal energy, 
${\cal E}$ and entropic energy, $-\T\calS$ and compare them to the counterpart 
quantifies from conventional KS calculations. Here $\calF_{el}$ is the 
electronic free energy, which conventionally is defined to be 
${\cal F}_{el} = \calF + E_{ion-ion} +  \int\, \dr v(\br) n(\br)$ with $\calF$ as defined in Eq.\ 
(\ref{eq:calFdefn}).

For T = 1 eV, LKTF performs best over $2.0 \le \rho \le 2.6$ g/cm$^3$.
However, this is a result of error cancellation.  
The LKTF pressure contribution from $E$ underestimates that from
conventional KS, while 
the entropic contribution does the opposite. 
For T = 5 eV, the  Perrot functional clearly works better.  
Even so, the thermodynamic contributions displayed in 
Fig.~\ref{fig:Si-Component-Perrot-5eV} show clearly that 
the comparatively good
performance is a consequence of error cancellation between 
contributions both of which are rather far from the conventional
KS values. Both cases shown also illustrate the underlying challenge: 
the conventional KS pressure is the
result of significant cancellation of the two thermodynamic 
contributions.

\begin{figure}[!htbp]
	\centering
	\includegraphics[width=1\linewidth]{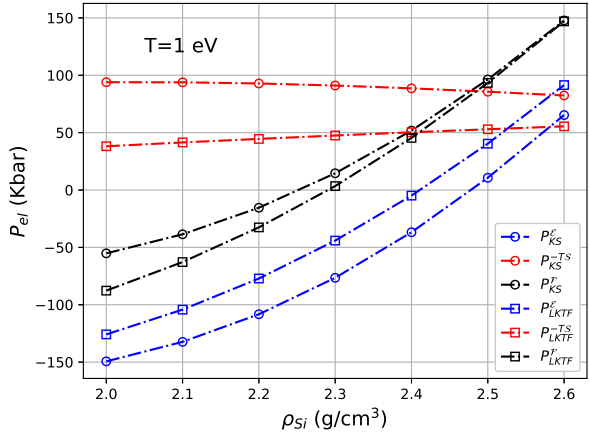}
	\caption[Si-LKT-Components]
	{ 
	Comparison of electronic pressure contributions from LKTF and 
conventional KS calculations for static \textrm{cd} Si at T = 1 eV. Superscript 
$\mathcal E$ denotes internal energy contribution, $\T{\mathcal S}$, the
 entropic contribution, and $\mathcal P$ the total.
	}
	\label{fig:Si-Component-1eV}
\end{figure}

\begin{figure}[!htbp]
	\centering
	\includegraphics[width=1\linewidth]{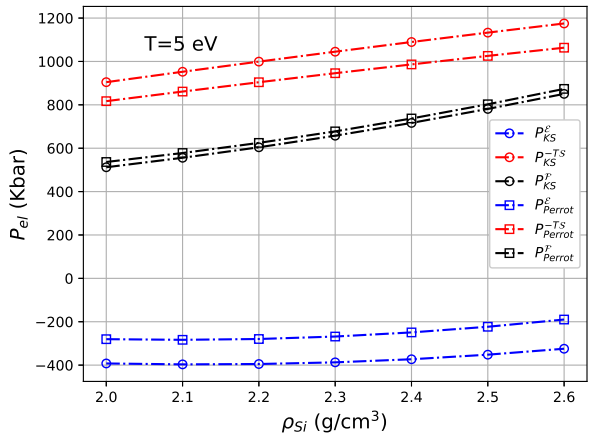}
	\caption[Si-Perrot-Components]
	{ 
	As in Fig.\ \ref{fig:Si-Component-1eV} for the Perrot 
functional versus conventional KS calculations at T=5 eV.
	}
	\label{fig:Si-Component-Perrot-5eV}
\end{figure}

\subsection{\textit{Ab initio} molecular dynamics}
\label{sub:aimd}

One of the strongest motivations for free energy OF-DFT is, as noted
already, the prospect of linear scaling of AIMD calculation costs with
respect to system size.  Thus we turn from static lattice EOS to AIMD
EOS calculations.

\begin{figure}[!htbp]
	\centering
	\includegraphics[width=\linewidth]{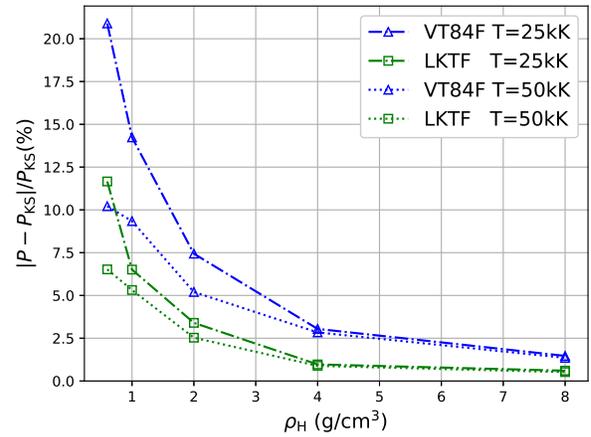}
	\caption[hydrogen density pressure]{Pressure error relative
to KS-AIMD 
as function of bulk density for H with LKTF (squares) and VT84F (triangles)
at T = 25 kK (dash-dotted curve) and 50 kK (dotted curve).
Densities are $0.6, 1.0, 2.0, 4.0, 8.0$ g/cm$^3$, 
	}
	\label{fig:hydrogen}
	\vspace*{-8pt}
\end{figure}

The EOS results for H from AIMD with the LKTF,  VT84F, and 
conventional KS-AIMD treatments are plotted in Fig.~\ref{fig:hydrogen}. 
For $\rho=0.6$ g/cm$^3$ and T = 25 kK, the relative error is reduced from 21\% for VT84F to 11\% for LKTF,
roughly a factor of two. As the temperature grows, the error from LKTF
decreases from 11\% to 6\%, while as the density increases, the relative
error rather quickly falls below 3.5\%.  This behavior is qualitatively
similar to what was found for VT84F \cite{Karasiev2013PRB}. The 
pressure error relative to conventional KS-AIMD results decreases 
as the density and/or the temperature increases.

\begin{table}[!htbp]
	\caption{H pressure at various densities and two
temperatures, T= 25 and 50 kK from AIMD simulations with  
 LKTF, VT84F, and conventional KS.  
After equilibration, pressures were averaged over 2000 steps. 
Andersen thermostat was used.}
	\centering
		\begin{tabular}{c|c|r r r r c}			
		\hline \hline
		T (kK) & $\rho_{\H}$ (g/cm$^3$)  & $P_{\rm KS}$ & $P_{\rm VT84F}$ & $P_{\rm LKTF}$ & (Mbar)\\
		\hline 
	     & 0.6 & 2.1 & 1.7 & 1.9 & \\
		 & 1.0 & 5.0 & 4.3 & 4.6 & \\
	25	 & 2.0 & 16.9 & 15.7 & 16.3 & \\
		 & 4.0 & 59.1 & 57.4 & 58.5 & \\
		 & 8.0 & 207.2 & 204.1 & 205.8 &\\
		\hline
	     & 0.6 & 3.9 & 3.5 & 3.6 & \\
		 & 1.0 & 8.0 & 7.2 & 7.5 & \\
	50	 & 2.0 & 22.7 & 21.5 & 22.2 & \\
		 & 4.0 & 70.6 & 68.6 & 69.9 & \\
		 & 8.0 & 229.5 & 226.5 & 228.3 & \\
		\hline
		\end{tabular}
	\label{tab:metals}
\end{table}

\begin{figure}
	\centering
	\includegraphics[width=\linewidth]
	{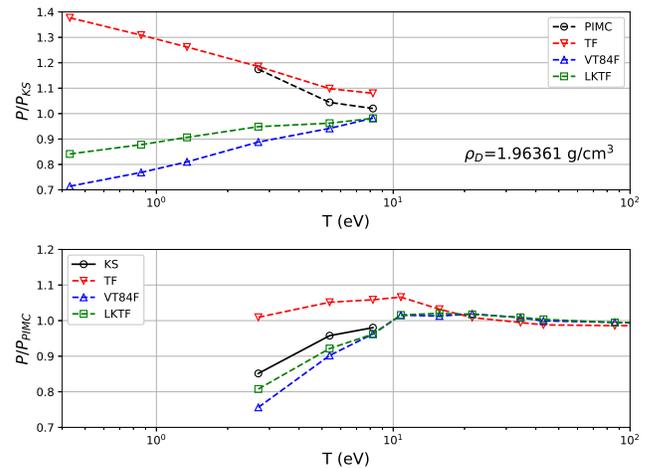}
	\caption[relative error for rs=1.4]{Relative pressures versus 
temperature for D at $\rho$=1.96361 ($r_s$ = 1.4 bohr) from PIMC, KS,  
LKTF, VT84F, and TF. Upper panel is relative to KS pressures, lower 
is relative to PIMC pressures.
	}
	\label{fig:rs1.4}
	\vspace*{-8pt}
\end{figure}

For D, we chose two bulk densities  $\rho_\D = $ 1.96361 ($r_s$ = 1.4
bohr), and  $\rho_\D$ = 4.04819 g/cm$^3$ ($r_s$ = 1.1 bohr) for which 
PIMC data are available \cite{Hu2011PRB}.  (Note that  data from SD$\beta$-vW14F calculations are unavailable for the lower 
density.)  We remark that
comparisons with the PIMC data involve the entire free energy
functional utilized. Hence those comparisons may be distorted by our use of 
a simple ground-state LSDA XC functional. That possible problem does not
arise in comparison with our KS-AIMD results, because those 
calculations used the same ground-state XC functional.  

For the lower
density, Fig.\ \ref{fig:rs1.4}
 displays the pressure as function of $T$ relative to both KS-AIMD
values ($P/P_{KS}$ and relative to PIMC results ($P/P_{PIMC}$).  
KS results are available
up to T = 95 350 K $\approx 8.2$ eV, while PIMC data are available
only for T $\ge$ 31 250 K, $\approx 2.7$ eV.  At the
lowest temperature, T=5 kK ($\approx 0.43$ eV), LKTF underestimates 
the pressure by $\approx  15$\%, while VT84F is worse, at about $30$\%.
TF, in contrast, drastically overestimates the low-T pressure by
almost $40$\%.  As T increases, the error from LKTF reduces 
quickly to an $\approx 5$\% underestimate at 31.25 kK  with continuing 
reduction as T increases. The T-dependence of $P/P_{KS}$ 
for VT84F is similar, but with about twice the error of LKTF.  
As a caution, note in the upper panel of the figure that the PIMC 
pressure at T = 31.25 kK deviates as much from
the KS pressure as does the TF pressure. We believe that this
deviation is a sign of well-known technical difficulties in PIMC
for comparatively low temperatures.  For T $\ge 100$ kK, however,
PIMC indisputably is a reliable reference.  In that regime
both LKTF and VT84F are reasonably accurate.  Both give pressures
that approach TF values (by construction) for large T.

\begin{figure}[!htbp]
	\centering
	\includegraphics[width=\linewidth]
	{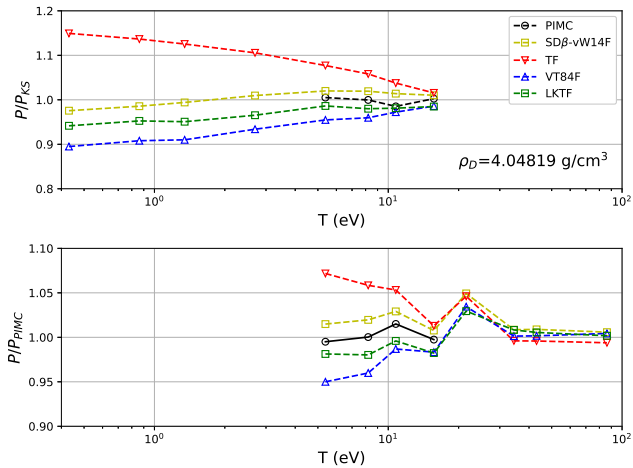}
	\caption[relative error for rs=1.1]{As in Fig. \ref{fig:rs1.4} for
D at $\rho$=4.04819 g/cm$^3$ ($r_s$=1.1 bohr) and with 
SD$\beta$-vW14F data as well.
	}
	\label{fig:rs1.1}
	\vspace*{-8pt}
\end{figure}

For the higher D density, Fig.\ \ref{fig:rs1.1} shows that the largest 
error relative to KS-AIMD pressure  still is at the lowest
temperature. LKTF underestimates the pressure by 7\% at most,
an error reduction of almost 2/3 compared to  VT84F. As in the
lower density case, TF again 
overestimates the low-T pressure, here by $\approx 14$\%.  Relative
to KS, the 
two-point functional, SD$\beta$-vW14F, achieves better performance
up to about T= 50 kK.  
Above that, LKTF is just as good.  Relative to the PIMC 
results, LKTF performs as well or better than SD$\beta$-vW14F.

\begin{figure}
	\centering
	\includegraphics[width=\linewidth]{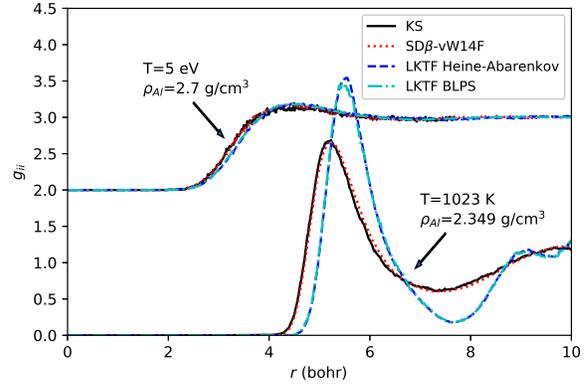}

	\caption[Al radial distribution function]{Al radial distribution 
function (RDF) from LKTF (blue dash), KS (black solid), 
and SD$\beta$-vW14F (red dotted) calculations for (a) T = 5 eV, 
$\rho =2.7$ g/cm$^3$ and  
(b) T = 1023 K, $\rho = 2.349$ g/cm$^3$.  The RDF for (b) is  
shifted upward by 2 for clarity of display.
	}
	\label{fig:Al}
	\vspace*{-8pt}
\end{figure}

For one further comparison, we also computed the radial distribution
function (RDF) of Al for two sets of state conditions: (a) T = 5 eV,
$\rho = 2.7$ g/cm$^3$, i.e, in the WDM regime; (b) T = 1023 K,
$\rho=2.349$ g/cm$^3$, i.e. near melting. Our calculations used the BLPS
local pseudo-potential, as before, as well as the Heine-Abarenkov 
\cite{Heine.Abarenkov.1964,Goodwin..Heine.1990} local pseudo-potential.  Fig.~(\ref{fig:Al}) displays
the results.  In the lower-T case, LKTF overestimates the height of
the first RDF peak relative to conventional KS value and shifts the peak position outward. This behavior is independent of detailed difference in the
local pseudopotential. The LPS and Heine-Abarenkov RDFs are virtually
indistinguishable.  Unsurprisingly,  SD$\beta$-vW14F does
much better, an obvious consequence of its intrinsic non-locality.   
For WDM conditions, LKTF 
delivers as good quality a RDF as the two-point functional
SD$\beta$-$\vW$14F.  Both are in good agreement with the conventional
KS RDF. This again is plausible because of the great reduction
in inhomogeneity upon going from T $\approx$ 0.09 eV to 5  eV.  

\section{Discussion and Summary}
\label{sec:discussion}

LKTF, the finite-T generalization of the LKT orbital-free kinetic energy
density functional presented here, represents a significant advance over
previously available one-point (semi-local) non-interacting free
energy functionals.  LKTF exploits non-universality in the form of
specific adaptation to near-nucleus properties of pseudo-densities. 
As a consequence, in both static lattice and AIMD calculations on
a few elemental systems, 
LKTF substantially reduces errors versus KS or KS-AIMD compared to the
previous best semi-local form, VT84F. Both of those constraint-based 
functionals deliver performance substantially superior to TF. 
At least for the Al RDF in the WDM regime, LKTF does as well as the
non-local SD$\beta$-vW14F. Wider usage of LKTF is needed both to
exploit its advantages and identify limitations.    

The improved performance of LKTF (relative to VT84F as the prior
benchmark) is obtained at least in part by
error cancellation between the kinetic and entropy contributions to
the non-interacting free energy.  Such cancellation may be system-dependent,
so reducing cancellation substantially while maintaining fidelity to 
conventional finite-T KS results is an important goal.
Two other matters of investigation are suggested by the LKTF performance.
One is whether the approximation of using Eq.\ (\ref{twominusF}) is 
inadequate and needs to be supplanted by solution of
the exact thermodynamic relation between $F_\sigma$ and $F_\tau$. Second
is whether the methodology of Ref. [\onlinecite{Karasiev2012PRB}] has some
unrecognized limitation that impacts the construction of functionals
such as VT84F and LKTF.

\section{Acknowledgments}
\label{sec:acknowledgments}
All the computations were performed on the Univ. Florida Research
Computing HiPerGator-II system. We thank Travis Sjostrom for
generously providing the SD$\beta$-$\vW$14F radial distribution
function data.  The majority of the work reported here was done while
KL was at Univ.\ Florida.  Both he and SBT were supported by
U.S.\ Dept.\ of Energy grant DE-SC 0002139.  VVK acknowledges support
by the Dept.\ of Energy National Nuclear Security Administration under
Award Number DE-NA0003856.

%

\end{document}